\documentclass[letterpaper,pra,twocolumn,showpacs,superscriptaddress]{revtex4}

\newif \ifpdf \ifx \pdfoutput \undefined \pdffalse
\else \pdfoutput=1 % we are running PDFLaTeX
\pdftrue \fi
\ifpdf
\usepackage[pdftex]{graphicx} \else \usepackage{graphicx} \fi
\usepackage{times}
\usepackage{graphicx}
\begin{document}
%%%%%%%%%%%%%%%%%%%%%%%%%%%%%%%%%%%%%%%%%%%%%%%%%%%%%%%%%%%%%%%
\title{Teleportation of continuous variable polarisation states}

\author{A. Doli\'{n}ska}
\affiliation{Quantum Optics Group, Department
of Physics, Faculty of Science, Australian National University, ACT
0200, Australia}
\author{B. C. Buchler}
\affiliation{Quantum Optics
Group, Department of Physics, Faculty of Science, Australian National
University, ACT 0200, Australia}
\author{W. P. Bowen}
\affiliation{Quantum Optics Group, Department of Physics, Faculty of
Science, Australian National University, ACT 0200, Australia}
\author{T. C. Ralph}
\affiliation{Centre for Quantum Computer Technology, Department of 
Physics,
University of Queensland, QLD 4072, Australia}
\author{P. K. Lam}
\affiliation{Quantum Optics Group,
Department of Physics, Faculty of Science, Australian National
University, ACT 0200, Australia}

\begin{abstract}
This paper discusses methods for the optical teleportation of
continuous variable polarisation states.  We show that using two pairs
of entangled beams, generated using four squeezed beams, perfect 
teleportation of
optical polarisation states can be performed.  Restricting ourselves
to 3 squeezed beams, we demonstrate that polarisation state
teleportation can still exceed the classical limit.  The 3-squeezer
schemes involve either the use of quantum non-demolition measurement
or biased entanglement generated from a single squeezed beam.  We
analyse the efficacies of these schemes in terms of fidelity, signal
transfer coefficients and quantum correlations.
\end{abstract}

 \date{\today} \maketitle
%%%%%%%%%%%%%%%%%%%%%%%%%%%%%%%%%%%%%%%%%%%%%%%%%%%%%%%%%%%%%%%
\section{Introduction}

Quantum teleportation \cite{ben93} is an important operation for the
transmission and manipulation of quantum states and information.  It
has been experimentally demonstrated in both discrete \cite{Discrete} 
and continuous variable \cite{Furusawa98.S,TelepLONG} regimes.  To 
date, continuous variable teleportation protocols have been performed 
solely on the quadrature
amplitudes of optical fields.  Recently there has been growing
interest in continuous variable polarisation states in the context of
quantum information schemes.  Experimental demonstrations of
polarisation squeezing \cite{bow021, Leuchs, Giacobino, Polzik,
Grangier} and entanglement \cite{bow022} have been performed. 
A practical advantage of polarisation states when applied to quantum
information networks is that a network-wide frequency reference is not
required \cite{Korolkova}.  Furthermore, quantum communication
networks are expected to require the ability to transfer quantum
information between optical and atomic states.  This has been
experimentally demonstrated between optical polarisation states and
atomic spin ensembles \cite{Polzik}.  It is then natural to ask how
quantum teleportation can be optimally implemented on continuous
variable polarisation states. 

This paper is arranged in the following way.  Section II reviews the
use of Stokes operators to characterise the quantum properties of
polarised light.  In Section III we discuss two commonly used
teleportation figures of merit in the context of quadrature
teleportation.  Section IV proposes a straightforward generalisation
of quadrature teleportation to polarisation teleportation, and
generalises the teleportation figures of merit to polarisation 
states. 
In Section V, VI and VII modifications of this protocol that optimise
these figures of merit are discussed.  We summarise and conclude in
Section VIII.

\section{Background}

\begin{figure}[b]
\begin{center}
\includegraphics[width=0.4\textwidth]{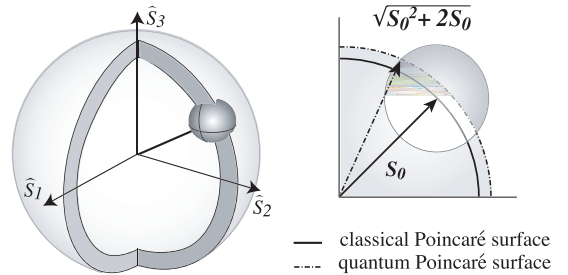}
    \caption{The quantum Poincar\'{e} sphere.  In the classical case
    $S_{0}$, the total photon number is the radius of the sphere;
    whereas in quantum picture the radius takes a larger value of
    $\sqrt{S_{0}^{2} +2S_{0}}$ due to the quantum uncertainty
    \cite{Korolkova}.  The presence of the uncertainty relations (eq.
    \ref{StokeUncertain}) manifests itself in the quantum noise
   ``ball'' as indicated.}
\label{SBall}
\end{center}
\end{figure}

In classical optics the polarisation state of light can be described
using Stokes parameters, where an arbitrary polarisation state is
decomposed into three components: linear (vertical/ horizontal),
diagonal (+45 / -45 degree) and circular (left/ right handed)
\cite{ClassStokes}.  This vector representation can be elegantly
visualised on a Poincar\'{e} sphere shown in FIG. \ref{SBall}.  The
orientation of the Stokes vector describes the polarisation state of
the laser beam with $\hat{S}_{1}$ giving the intensity difference
between the horizontally and vertically polarized components of the
beam and $\hat{S}_{2}$ giving the intensity difference between the
diagonally and anti-diagonally polarized components.  The azimuthal
deviation from the $\hat{S}_{1}-\hat{S}_{2}$ plane towards the
$\hat{S}_{3}$ axis indicates the ellipticity of the polarization
state.
By drawing an analogy with classical Stokes parameters a set of
Stokes operators can be defined, providing a convenient
description of the quantum polarisation properties of light:
\cite{Robson,Electron}
%%% STOKES
\begin{eqnarray}
\hat{S}_{0} & = &\hat{a}_{H}^{\dagger}\hat{a}_{H} +\hat
{a}_{V}^{\dagger}\hat{a}_{V} = \hat{n}_{H} +\hat{n}_{V}, \nonumber \\
\hat{S}_{1} & = &\hat{a}_{H}^{\dagger}\hat{a}_{H} -\hat
{a}_{V}^{\dagger}\hat{a}_{V} = \hat{n}_{H} -\hat{n}_{V}, \nonumber \\
\hat{S}_{2} &=&\hat{a}_{H}^{\dagger}\hat{a}_{V}e^{i\theta} +\hat
{a}_{V}^{\dagger}\hat{a}_{H}e^{-i\theta} = \hat{n}_{D}
-\hat{n}_{\bar{D}}, \nonumber \\
\hat{S}_{3} &=& i\hat{a}_{V}^{\dagger}\hat{a}_{H}e^{-i\theta} -i\hat
{a}_{H}^{\dagger}\hat{a}_{V}e^{i\theta} \label{TH} = \hat{n}_{R}
-\hat{n}_{L}.
\label{Stokes}
\end{eqnarray}
Here the polarisation mode is constructed in terms of annihilation,
$\hat{a}$, and creation, $\hat{a}^{\dagger}$, operators of the
horizontal (H) and vertical (V) constituent modes, with a phase, 
$\theta$, between them.  These operators
can be written as $\hat{a}(t) = \alpha + \delta\hat{a}(t)$, where
$\alpha$ is the classical amplitude and $\delta\hat{a}(t)$ is the
operator containing the quantum fluctuations with $[\delta\hat{a}(t),
\delta \hat{a}^{\dagger}(t)] = 1 $ and 
$\langle\delta\hat{a}(t)\rangle = 0$.  We will assume that $|\alpha| 
= \langle |\hat{a}(t)|\rangle \gg 
\langle|\delta\hat{a}(t)|^{2}\rangle$ allowing a linearisation of the 
operator equations.

$\hat{S}_{0}$ commutes with the other Stokes operators and its
expectation value is proportional to the total intensity of the light
beam.  $\hat{S}_{1}$, $\hat{S}_{2}$ and $\hat{S}_{3}$, however, obey a
coupled set of commutation relations and are isomorphic to the Pauli
matrices: $[\hat{S}_{l}, \hat{S}_{m}] = 2i \hat{S}_{n}$, where $\{l,
m, n\} = \{1,2,3\}$ and are cyclically interchangeable.  This says
that simultaneous measurements of these Stokes operators are, in 
general,
impossible and their variances are restricted by
\begin{equation}
  V_{l} V_{m} \geq |\langle \hat{S}_{n}\rangle|^{2} 
\label{StokeUncertain}
\end{equation}
Here, $V_{l}=\langle (\hat{S}_{l})^{2}\rangle - \langle
\hat{S}_{l}\rangle^{2}$ is the variance of each Stokes operator.

The length of the quantum Stokes vector in FIG.~\ref{SBall} is
$\sqrt{\langle S_{0}^{2}\rangle +2\langle S_{0} \rangle}$, which
always exceeds its classical counterpart.  The coupled uncertainty
relations of the Stokes variances in equation (\ref{StokeUncertain})
are exhibited further in the appearance of a three dimensional noise
``ball'', superimposed on the Poincar\'{e} surface, at the end of the
Stokes vector.  In the case of coherent polarisation states this ball
is spherical.

The field operators $\hat{a}(t), \hat{a}^{\dagger}(t)$ are now 
expanded in terms of their DC and fluctuating componants.  Keeping 
only the first order fluctuation terms,  Eq. (\ref{Stokes}) yield 
linearised equations for the fluctuations in the Stokes operators:
%linearised Stokes
\begin{eqnarray}
\delta\hat{S}_{0} &=& \alpha_{H} \delta X^{+}_{H} +\
\alpha_{V}\delta X^{+}_{V},
\label{st0} \\
\delta\hat{S}_{1} &=& \alpha_{H} \delta X^{+}_{H} -\
\alpha_{V}\delta X^{+}_{V},
\label{st1} \\
\delta\hat{S}_{2} &=& \alpha_{H} (\delta X^{-}_{V}
\sin \theta +\delta X^{+}_{V}
 \cos\theta ) + \nonumber \\
&& \alpha_{V}\
(\delta X^{+}_{H}\
\cos \theta -\delta X^{-}_{H} \sin
\theta),
\label{st2} \\
\delta\hat{S}_{3} &=& \alpha_{H} (\delta X^{+}_{V}
 \sin \theta -\delta X^{-}_{V}
 \cos \theta ) + \nonumber \\
&& \alpha_{V}\
(\delta X^{-}_{H} \cos
\theta +\delta X^{+}_{H} \sin \theta),
\label{st3}
\end{eqnarray}
where $\hat{X}^{+(-)}$ are the usual amplitude (phase) quadrature
operators, defined as $\delta\hat{X}^{+} = (\delta\hat{a} +
\delta\hat{a}^{\dagger})$, and $\delta\hat{ X}^{-} = i
(\delta\hat{a}^{\dagger} - \delta\hat{a})$.  It can be seen from
equations (\ref{st1}-\ref{st3}) that the linearised Stokes operators
are a linear combination of the quadrature operators for the two
modes.

In this paper we are interested in fluctuations at a frequency
$\omega$ around the optical carrier frequency.  The Fourier
transform of the time domain Stokes operators will be taken from
now on, with all the operators being in the frequency domain.  We
include the signal at frequency $\omega$, encoded on polarisation
modulation as a classical fluctuations term, making $\hat{a} =
\alpha_{c} + \delta\hat{a}_{q} + \delta a_{c}$.  Unlike quantum
fluctuations $\delta\hat{a}_{q}$, the introduced $\delta a_{c}$ term
is purely classical with [$\delta a_{c}, \delta a_{c}^{\dagger}$] = 
0. 
The $\hat{a}$ operator expansions substituted into Stokes equations
(\ref{Stokes}) yield linearised equations (\ref{st1}-\ref{st3}) in
frequency domain where $\delta\hat{X}^{\pm} =\delta X^{\pm}_{c}
+\delta \hat{X}^{\pm}_{q}$.  Hence there are two independent sources
of fluctuations, the classical signal ($c$) and the quantum noise
($q$).  The variances, $V(\delta\hat{S}_{l})$, of the Stokes operators
may be calculated from equations (\ref{st1}-\ref{st3}).
%  ************************** Variances for Stokes 
\begin{widetext}
\begin{eqnarray}
V_{S1} & = & \alpha_{H}^{2} (V_{H,c}^{+}+V_{H,q}^{+}) + \alpha_{V}^{2}
(V_{V,c}^{+} + V_{V,q}^{+}) +\ 2\alpha_H \alpha_V\ \langle 
\delta X^+_{V,c} \delta X^+_{H,c} \rangle, \label{var1} \\
V_{S2} & = & \alpha_{H}^{2}{(\cos \theta )^2}(V_{V,c}^{+} +
V_{V,q}^{+}) + \alpha_{V}^{2}{{(\cos \theta)}^2} (V_{H,c}^{+}+
V_{H,q}^{+}) + \alpha_{H}^{2} {(\sin \theta )^2} (V_{V,c}^{-}+
V_{V,q}^{-}) + \alpha_{V}^{2}{{(\sin \theta )}^2} (V_{H, c}^{-}+ 
V_{H,q}^{-})
\nonumber \\
&& + \ 2\alpha _H\alpha _V\sin\theta\ \cos\theta \ \langle 
\delta X^-_{V,c}\delta X^+_{H,c}\rangle \ 
+ \ 2\alpha _H\alpha _V (\cos \theta )^2 \langle
\delta X^+_{V,c}\delta X^+_{H,c}\rangle \ 
+ \ 2\alpha _H^2\sin\theta\ \cos\theta \ \langle 
\delta X^+_{V,c}\delta X^-_{V,c} \rangle \nonumber \\
&& - \ 2\alpha _H\alpha _V\sin\theta\ \cos\theta \ \langle
\delta X^+_{V,c}\delta X^-_{H,c}\rangle 
-\ 2\alpha _H\alpha _V (\sin \theta )^2 \langle
\delta X^-_{V,c}\delta X^-_{H,c}\rangle
-\ 2\alpha_V^2\sin\theta\ \cos\theta \ \langle 
\delta X^+_{H,c} \delta X^-_{H,c} \rangle
\label{var2} \\
V_{S3} & = & \alpha_{H}^{2} {{(\cos \theta )}^{2 }} (V_{V,c}^{-}+
V_{V,q}^{-})  + \alpha_{V}^{2}{{(\cos \theta )}^2} (V_{H,c}^{-}+
V_{H,q}^{-})  + \alpha_{H}^{2} {{(\sin \theta )}^{2 }} (V_{V,c}^{+}+
V_{V,q}^{+})  + \alpha_{V}^{2}{{(\sin \theta)}^2} (V_{H,c}^{+}+ 
V_{H,q}^{+})
\nonumber \\
&& + 2\alpha _H\alpha _V\sin\theta\ \cos\theta \ \langle 
\delta X^+_{V,c} \delta X^-_{H,c} \rangle \ 
+  \ 2\alpha _H\alpha _V (\sin \theta )^2 \langle 
\delta X^+_{V,c} \delta X^+_{H,c} \rangle  
+ \ 2 \alpha_V^2 \sin\theta\ \cos\theta \ \langle \delta X^+_{H,c}
\delta X^-_{H,c} \rangle \nonumber \\
&& - \ 2\alpha _H\alpha _V\sin\theta\ \cos\theta \ \langle
\delta X^-_{V,c} \delta X^+_{H,c} \rangle  
- 2\alpha _H\alpha _V (\cos \theta )^2 \langle 
\delta X^-_{V,c} \delta X^-_{H,c} \rangle \ 
- 2\alpha _H^2\sin\theta\ \cos\theta \ \langle 
 \delta X^+_{V,c} \delta X^-_{H,c}
\rangle 
\label{var3}
\end{eqnarray}
\end{widetext}
%******************************************************************************

The variance terms with subscript 'c' represent a delibrately applied
signal, distinct to the quantum noise terms with subscript 'q'.  In
general, classical modulation correlations can exist and additional
cross terms, such as $\langle \delta X^{+}_{H,c} \delta
X^{+}_{V,c}\rangle$, may appear.  These are included for completeness,
although they are not considered in the modelling that follows in
later sections.  In the following sections, we will assume the light
beams are pure states with Gaussian statistics.  Unless squeezed, the
quantum terms will be at the standard quantum limit and
$V^{\pm}_{H/V,q}$ = 1.

\section{Figures of Merit for Quadrature Teleportation}

The figures of merit that we consider here for polarisation
teleportation are generalisations of those previously used for
quadrature teleportation, namely the T-V measure and fidelity
\cite{TelepLONG}.  In this section, we present the relevant
definitions of quadrature teleportation.  The extension of the
parameters is then presented in later sections.

Fidelity is one way to quantify the success of a quantum state
reconstruction for many quantum protocols.  It is given by the overlap
integral of the initial and final wave-functions, $\mathcal{F} =
|\langle \psi_{\rm in} |\hat{\rho}_{\rm out}|\psi_{\rm in}
\rangle|^{2}$, where $|\psi_{\rm in} \rangle$ is the input state, and
$\hat{\rho}_{\rm out}$ is the density operator of the output.  For
Gaussian input states the statistics of a laser beam are fully
described by the first two statistical moments: the mean and the
variance.  When unity gain is assumed for the reconstruction, that is,
the output state has the same classical amplitude as the input, and
when the input states are coherent, i.e. $V^{\pm}_{\rm in} =1$, the
expression for fidelity is given by
\begin{equation}
\mathcal{F} = \frac{2}{\sqrt{(V^{+}_{\rm out} +1) (V^{-}_{\rm out}
+1)}}.
\label{fidelity}
\end{equation}
$V^{\pm}_{\rm out}$ are the output quadrature variances.  Variations
away from unity gain typically lead to an exponentially decreasing fidelity
value \cite{TelepLONG}.

The case of $\mathcal{F}$ = 0 implies the input and output are
orthogonal and bare no resemblance to each other, while $\mathcal{F}
=$ 1 suggests perfect reconstruction of the input.  In the absence of
entanglement, the fidelity limit for the quadrature teleportation of 
a coherent state  is $\mathcal{F} \leq \frac{1}{2}$ 
\cite{Furusawa98.S}.  

Another useful way of quantifying teleportation is via a T-V diagram
\cite{Teleportation1}.  Here two parameters are considered.  The first
parameter is the signal transfer coefficient $T^{\pm}$, which is the
ratio of the signal-to-noise ratio $\mathcal{R}$ of the output to that
of the input for a given quadrature,
\begin{equation}
T_{q} = T^{+} + T^{-} = \frac{\mathcal{R}^{+}_{\rm
out}}{\mathcal{R}^{+}_{\rm in}}+ \frac{\mathcal{R}^{-}_{\rm
out}}{\mathcal{R}^{-}_{\rm in}}
\label{Tq}
\end{equation}
When no information is recovered there is no signal, hence $T_{q}=$ 0.
For ideal teleportation, the transfer coefficient has 
$\mathcal{R}^{\pm}_{\rm
in} = \mathcal{R}^{\pm}_{\rm out}$ for both quadratures, as the 
vacuum noise
problem is circumvented.  This gives the ideal two quadrature limit
of $T^{\rm max}_{q} =$ 2.  The classical limit at unity gain is given
by $T^{\rm classical}_{q} =\frac{2}{3}$.  
% As in the classically
% limited fidelity case, this limit is due to the double noise penalty
% which couples a unit of uncorrelated vacuum to the signal when
% measuring and when reconstructing the state.

The second parameter of the T-V diagram is the conditional variance,
$V_{\rm cv}= \frac{1}{2}(V^{+}_{\rm cv}+ V^{-}_{\rm cv})$, which is a
measure of the correlation between the input and the output
quadratures, and is defined as
\begin{equation}
V^{\pm}_{\rm cv} = V^{\pm}_{\rm out} - \frac{\langle |
\delta\hat{X}^{\pm}_{\rm in}\delta \hat{X}^{\pm}_{\rm out}|^{2} 
\rangle
}{V^{\pm}_{\rm in}}.
\label{CV}
\end{equation}
For Gaussian input states, it can be shown that $V^{\pm}_{\rm cv} =
V^{\pm}_{\rm out} (1 - T^{\pm})$, where $V_{\rm out}$ is the output of
the system with no signal input \cite{Teleportation1}.  The
conditional variance is a measure of quantum correlation between the
input and the output states and it reflects the amount of noise added
to the output state by the teleporter.  Ideal quadrature teleportation
replicates the input exactly, giving the lower bound of $V_{\rm
cv}^{\rm min} = 0$.  At unity gain the classical limit is again the
double vacuum noise penalty.  Hence $V_{\rm cv}^{\rm classic} = 2$.

The $T_{q}$ and $V_{\rm cv}$ parameters can be plotted on a T-V
diagram as a function of the teleportation feed-forward gains.  Once
evaluated, both equation (\ref{Tq}) and (\ref{CV}) become independent
of the input signal amplitudes and equation (\ref{CV}) is also 
independent of the input noise.

\section{Polarisation state teleportation with twin teleporters}
\label{Twins}

\begin{figure}[ht]
\centering
\includegraphics[width=0.4\textwidth]{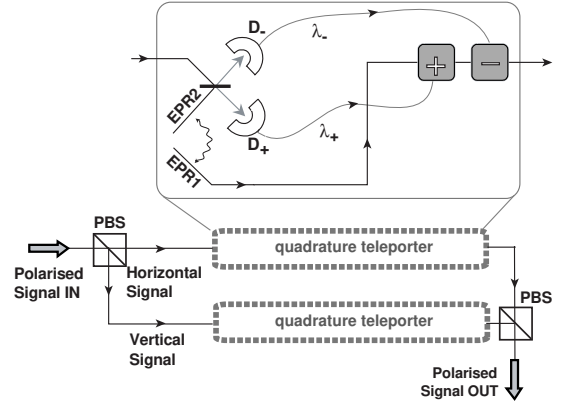}
     \caption{Polarisation state teleportation scheme with twin
     teleporters.  EPR1,2: two entangled beams; $D_{\pm}$:
     amplitude/phase homodyne detectors; +/- : amplitude/ phase
     modulators; $\lambda_{\pm}$: amplitude/ phase feedforward gains;
     PBS: polarising beamsplitter.  A standard quadrature teleporter
     is shown in the insert.}
\label{4sqTelExp}
\end{figure}

We note from equations (\ref{var1}-\ref{var3}) that polarisation
states can be completely described by the quadrature amplitudes of
both the horizontal and vertical polarisation modes.  The obvious way 
to teleport an input polarisation state is, therefore, to decompose 
the input beam into a horizontally and a vertically polarised beam 
via a polarising beamsplitter as shown in FIG.~\ref{4sqTelExp}
Two standard continuous variable quadrature amplitude teleporters, one
for each polarisation mode, can be used to teleport the orthogonally
polarised beams The complete task thus requires four squeezed beams
for the generation of two pairs of quadrature entanglement.  Finally,
the teleported states are recombined at the receiving station using
another polarising beamsplitter.

The teleportation fidelity for this system is shown in
FIG.~\ref{SQDFidel}(a).  Assuming that all 4 beams are equally
squeezed, the expression for the fidelity of the twin teleporters
scheme becomes,
\begin{equation}
     \mathcal{F} = \frac{1}{(V_{\rm SQ}+1)^{2}},
\label{FidelitySQD}
\end{equation}
where $V_{\rm SQ}$ is the variance of the squeezed quadratures of the
beams used to produce the entanglement.  Since the fidelities for the
vertical and horizontal modes are independent, the fidelity is
calculated from a four dimensional overlap integral between the input
and output states.  Equation (\ref{FidelitySQD}) is derived simply
by squaring the quadrature teleporter fidelity.  We note that
the classical limit of this polarisation teleporter is $\mathcal{F} 
\leq
\frac{1}{4}$ and ideal polarisation teleportation has fidelity 1. 

\begin{figure}[ht]
\centering
\includegraphics[width=0.4\textwidth]{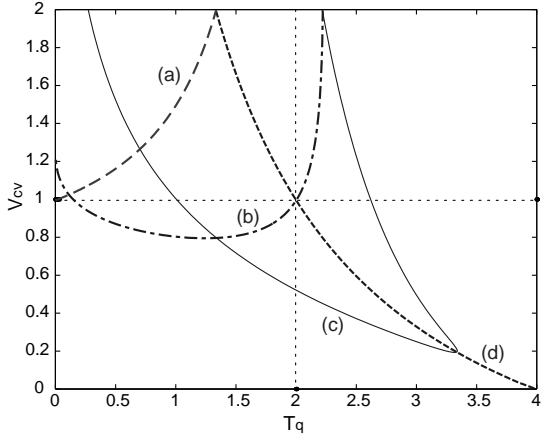}
     \caption{T-V plot of the teleportation of polarisation state with
     twin teleporters (a) with coherent states (b) with 3 dB squeezing
     ($V_{\rm SQ} = 0.5$), (c) with 10 dB squeezing ($V_{\rm SQ} =
     0.1$) as a function of feedforward gain.  (d) Locus of unity gain
     points from no squeezing to perfect squeezing.}
\label{4sqTV}
\end{figure}

The results of T-V analysis for this scheme are illustrated in
FIG.~\ref{4sqTV}.  Similar to the quadrature teleporter, the
conditional variance is now extended to $V_{\rm cv} =
\frac{1}{4}(V^{+}_{\rm cv,H}+ V^{-}_{\rm cv,H}+ V^{+}_{\rm cv,V}+
V^{-}_{\rm cv,V})$ and the total signal transfer coefficient is now
given by $T_{q} = T^{+}_{H}+ T^{-}_{H}+ T^{+}_{V}+ T^{-}_{V}$.  For
ideal squeezing, we obtain $V_{\rm cv} \rightarrow 0$ and $T_{q}
\rightarrow 4$.

\begin{figure}[ht]
\begin{center}
\includegraphics[width=0.47\textwidth]{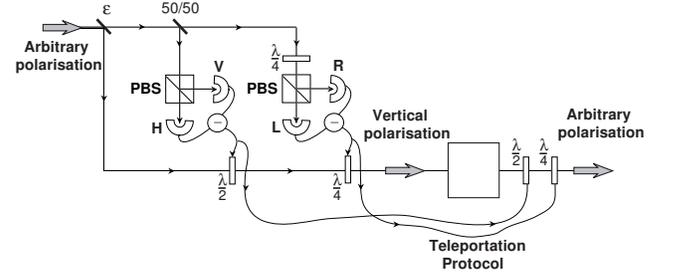}
    \caption{Classical control system for measuring and controlling
    the polarisation of carrier field.  This figure demonstrates that
    input polarisation state can be measured and fed forward to
    control the polarisation state of another beam.  $\varepsilon$:
    beamsplitter with low transmittivity, H/V: horizontal/ vertical
    polarisation detection, R/L: right/ left circular polarisation
    detection, PBS: polarising beamsplitter, $\frac{\lambda}{2}$:
    half-wave plate, $\frac{\lambda}{4}$: quarter-wave plate.  The
    vertical output is subsequently teleported by a chosen protocol
    and returned back to its original polarisation at the receiving
    station.  }
\label{Control}
\end{center}
\end{figure}

So far, we have chosen to ignore the classical amplitude of our input
state.  Although the fluctuations in the input polarisation are
teleported by the twin teleporters, the polarisation of the input
carrier field is not teleported.  This is, at first thought, analogous
to quadrature teleportation where the carrier amplitude, or the 
optical
intensity, of the input beam is assumed to be unimportant in the
reconstruction of the quantum state at the sideband frequency.  
Besides,
it is relatively trivial to replicate the input intensity at the
output.  Interestingly however, equations (\ref{var1}-\ref{var3})
suggest that the polarisation of the input carrier field cannot be
ignored in the teleportation of polarisation states.  This is due to
the fact that uncertainty relations of Stokes operators are directly
scaled by the carrier polarisation.  The carrier field polarisation
consists of two amplitudes (the horizontal and vertical components) as
well as one relative phase angle.  Polarisation fluctuations will only
be teleported properly provided the input polarisation is known and
the output polarisation is set to be identical.  A complete
polarisation teleporter would therefore include the twin teleporters
plus an optical setup presented in FIG.~\ref{Control} to shift an
arbitrary carrier field polarisation to a set polarisation and then,
after the teleportation protocol, return it to its original
polarisation.

\section{SQD-teleporter scheme}
\label{SQDtelSchm}

Inspection of the Stokes operators shows that since $\hat{S_{0}}$
commutes with $\hat{S_{1}}$, $\hat{S_{2}}$ and $\hat{S_{3}}$, it can
be measured with no penalty on the remaining three operators.  For
quadrature teleportation two squeezed beams enable teleportation of
two variables, $\delta\hat X^{-}$ and $\delta\hat X^{+}$.  This raises
the question of whether polarisation teleportation could be achieved
using only three squeezed beams (for $\hat{S_{1}}$, $\hat{S_{2}}$ and
$\hat{S_{3}}$) rather than the four utilised in the previous scheme 
\footnote{This can be done without loss of generality so long as the
setup in FIG.~\ref{Control} is utilised.}. Choosing the polarisation of
the carrier beam to be vertical, causes the $\alpha_{H}$ terms in
equations (\ref{st1}-\ref{st3}) to vanish, giving
\begin{eqnarray}
\delta\hat{S}_{1}^{\rm SQD}& = & - \alpha _{V}\delta X^{+}_{V},
\label{Lst1} \\
\delta\hat{S}_{2}^{\rm SQD}& = & \alpha _{V}(\delta X^{+}_{H} \cos
\theta -\delta X^{-}_{H} \sin \theta ), \label{Lst2} \\
\delta\hat{S}_{3}^{\rm SQD}& = &\alpha _{V}(\delta X^{-}_{H} \cos
\theta +\delta X^{+}_{H} \sin \theta ).
     \label{Lst3}
\end{eqnarray}

The linearised variances for the vertical carrier Stokes fluctuations
from equations (\ref{var1}-\ref{var3}) now simplify to
\begin{eqnarray}
V_{\delta\hat{S}_{1}}^{\rm SQD} & = & \alpha _{V}^{2} V_{V}^{+},
\label{Lvar1}  \\
V_{\delta\hat{S}_{2}}^{\rm SQD} & = & \alpha _{V}^{2}(\cos \theta 
)^{2}\
V_{H}^{+}  +  \alpha _{V}^{2}(\sin \theta )^{2}\
V_{H}^{-} -  \\ \nonumber
&& 2 \alpha_{V}^{2}\sin \theta \cos \theta  \langle
\delta X^{+}_{H,c} \delta X^{-}_{H,c}\rangle,
\label{Lvar2}  \\
V_{\delta\hat{S}_{3}}^{\rm SQD} & = & \alpha _{V}^{2}(\cos \theta 
)^{2}\
V_{H}^{-}  +  \alpha _{V}^{2}(\sin \theta )^{2}\
V_{H}^{+} +  \\ \nonumber
&& 2 \alpha_{V}^{2}\sin \theta \cos \theta  \langle
\delta X^{+}_{H,c} \delta X^{-}_{H,c}\rangle.
\label{Lvar3}
\end{eqnarray}
where the variances $V_{H/V}^{\pm} = V_{\rm c}^{\pm} + V_{\rm
q}^{\pm}$, are the sum of classical signal and quantum fluctuation
variances.  The classical cross correlation terms in equations
(\ref{var1}-\ref{var3}) have now reduced so that only correlations
between the phase and amplitude quadratures of the horizontal input
mode remain.

The phase angle $\theta$ has no affect on the classical polarisation
since $\alpha_H=0$, therefore making the angle between $\alpha_V$ and
$\alpha_H$ meaningless.  It does nevertheless appear in the
expressions for the variances of the Stokes operators, although the
angle is not referenced to a coherent field.  The situation is
analogous to the case of a squeezed vacuum state where its quadrature
angle, although lacking reference to a coherent amplitude, affects 
the variance.

The uncertainty relations in Eq.~(\ref{StokeUncertain}) are strongly
affected by choosing $\langle\hat{a}_{H}\rangle = 0$ since this also
implies $\langle \hat{S}_{2}\rangle=\langle \hat{S}_{3}\rangle=0$. 
>From Eq.~(\ref{StokeUncertain}), the only uncertainty remaining is
that between $\hat{S}_{2}$ and $\hat{S}_{3}$.  Quantum teleportation
of these two quantities can be achieved via a single entangled pair. 
$\hat{S}_{1}$ on the other hand commutes with $\hat{S}_{2}$ and
$\hat{S}_{3}$ and can be determined without disturbing them, therefore
its reconstruction does not require a second entangled pair.  In other
words, equations (\ref{Lst2}) and (\ref{Lst3}) are seen to completely
decouple from equation (\ref{Lst1}).  The vertical amplitude
fluctuations of $\delta\hat{S}_{1}^{\rm SQD}$ can therefore be
reproduced by a single quadrature (SQD) measurement \cite{SQDsqz1}.

\begin{figure}[ht]
\centering
\includegraphics[width=0.4\textwidth]{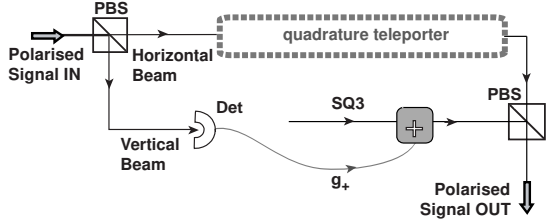}
     \caption{SQD-teleporter experimental setup consisting of a direct
     detection SQD measurement and a quadrature teleporter circuit.  
Det:
     standard amplitude detector, + : amplitude modulator, $g_{+}$:
     amplitude modulator gain, PBS: polarising beamsplitter.  }
\label{SQDTelExp}
\end{figure}

The schematic of this SQD protocol is shown in FIG.~\ref{SQDTelExp}.
Vertically polarised light is incident at the input polarising 
beamsplitter. The bright vertical light mode is reflected and 
detected. The resulting photocurrent is used to control the amplitude 
modulation of a vertically polarised squeezed beam, SQ3.  The 
amplitude quadrature of the modulated beam SQ3 will, in the limit of 
ideal squeezing and appropriate feed-forward gain, be identical to 
$\delta\hat{X}_{V}^{+}$, the amplitude quadrature of the the 
vertically polarised light at the input to the teleporter.  Since 
$\delta\hat{S}_{1}^{\rm SQD}\propto \delta\hat{X}_{V}^{+}$ this 
single quadrature feed-forward loop is enough to teleport 
$\delta\hat{S}_{1}$
 The quadrature teleportation protocol, using an EPR pair, transfers 
the fluctuations of $\delta\hat{S}_{2}^{\rm SQD}$ and 
$\delta\hat{S}_{3}^{\rm SQD}$ onto
the horizontally polarised output beam EPR1 \cite{Teleportation1}. 
The vertical and horizontal output modes are then combined via a
second polarising beamsplitter and the polarisation information is
recreated.  It is important to ensure that the horizontal output mode
(EPR1) has much less power than the vertical output beam SQ3, in order
to preserve the input polarisation.

The above scheme is not necessarily limited only to vertically
polarised input states.  An arbitrary input state can be rotated using
a variable half and quarter-wave plate arrangement and feedback loops,
such as that in FIG.~\ref{Control}, to ensure all of the light power
is in the SQD part of the system and its polarisation is vertical.
Once the protocol is complete, it can be rotated back to its original
polarisation.

The amplitude squeezing of SQ3 enables, in theory, a perfect
reproduction of the single amplitude quadrature fluctuation
$\delta\hat{X}_{V}^{+}$.  The complete polarisation teleportation
system now uses only an entangled pair and one additional squeezed
beam.

%%%%%% TV %%%%%%%%
An interesting characteristic of the measurement of the vertical
polarization is that the signal transfer is best in the limit of
infinite gain.  On the other hand, the conditional variance of the
vertical polarisation cannot be improved as there are no correlations
between the detected beam and the squeezed reconstruction beam.

It is possible however, to represent the entire system on a single T-V
diagram with $T_{q} = T^{+}_{H} + T^{-}_{H} + T^{+}_{V}$ and $V_{\rm
cv} = \frac{1}{3}(V^{+}_{\rm cv,H} + V^{-}_{\rm cv,H} + V^{+}_{\rm
cv,V})$.  Since the phase quadrature $\delta\hat{X}^{-}_{V}$ is
irrelevant to the polarisation description of the state (equations
(\ref{Lvar1}-\ref{Lvar3})), it is reasonable not to include it in the
T-V analysis, which relates to the polarisation information
transferred during the teleportation process.

\begin{figure}[ht]
\centering
\includegraphics[width=0.4\textwidth]{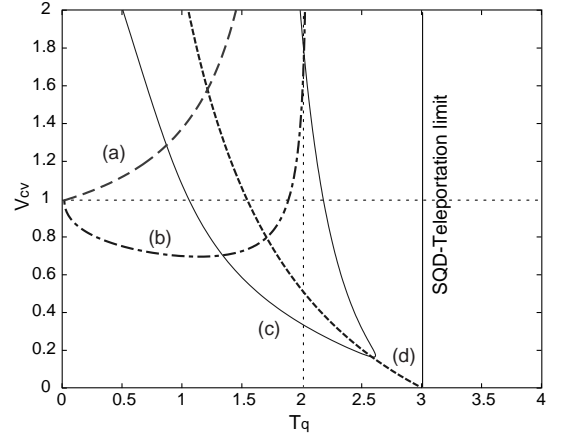}
     \caption{The SQD-teleporter T-V plot using only three quadratures
     of interest, with $V_{SQ3}=V_{SQ}$.  (a) with coherent states
     (b) with 3 dB squeezing ($V_{\rm SQ} = 0.5$), (c) with 10 dB
     squeezing ($V_{\rm SQ} = 0.1$) and (d) Locus of unity gain points
     from no squeezing to perfect squeezing }
\label{SQDTV}
\end{figure}

For simplicity, the choice of $V_{SQ3} = V_{SQ}$ is made for the
remainder of this section.  FIG.~\ref{SQDTV} shows a resulting three
quadrature ($X_{H}^{+},\ X_{H}^{-}, X_{V}^{+} $) T-V plot.  For ideal
squeezing of all three beams the minimum conditional variance $V_{\rm
cv}\rightarrow 0$, and the maximum signal transfer coefficient $T_{q}
\rightarrow 3$, can still be reached.

\begin{figure}[ht]
\centering
\includegraphics[width=0.4\textwidth]{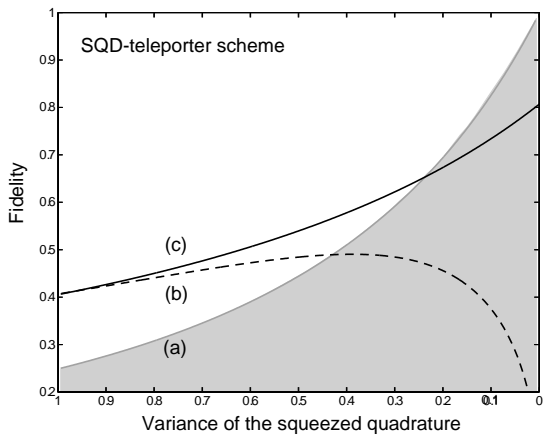}
     \caption{The fidelity curves for (a) the twin teleporter system;
     (b) the SQD-teleporter system with $V_{\rm SQ3} = V_{\rm SQ}$,
     (c) the SQD-teleporter system with no squeezing on the SQ3 beam}
\label{SQDFidel}
\end{figure}

%%%%%% Fidelity %%%%%%%
In the SQD scheme, beam SQ3 needs to be amplitude squeezed in order to
reduce any noise in the signal quadrature.  As a result, the phase
quadrature becomes very noisy as ${1}/{V_{SQ3}} \rightarrow \infty$. 
The unfavourable consequence of this is that the fidelity of the
scheme is found to be vanishingly small.  The fidelity equation
(\ref{fidelity}) reduces to
\begin{equation}
\mathcal{F}_{\rm SQD} = \frac{2}{(1+V_{\rm SQ})\sqrt{(V_{\rm SQ3} +2)\
(\frac{1}{V_{\rm SQ3}} +1)}}
\label{fidelity-SQD}
\end{equation}
In fact, the maximum fidelity of the replicated quantum state is
$\mathcal{F}$ = $\sqrt{(2/3)}$, attained when there is no SQ3
squeezing at all.  FIG.~\ref{SQDFidel} shows two possible fidelity
curves, with and without SQ3 beam squeezing.  The SQD-teleporter curve
exceeds the results of the twin teleporters for all squeezing levels
up to 80$\%$, even though less resources are used.  This is because 
there are fewer measurement penalties in the 3 beam case.  When 
performing classical teleportation (i.e. teleportation with coherent 
states in place of entanglement) of all 4 quadratures, each 
quadrature reconstruction will degrade the fidelity.  Classical 
teleportation of only 3 quadratures means the fourth quadrature is 
not degraded and therefore does not contribute to reducing the 
fidelity. The classical limit in the case of the SQD protocol may 
then be redefined by substituting $V_{\rm SQ}=V_{\rm SQ3}=1$ in Eq. 
\ref{fidelity-SQD}, giving $\mathcal{F}=1/\sqrt{6}$.

\section{Biased entanglement teleporter scheme}

\begin{figure}[b]
     \centering
     \includegraphics[width=0.4\textwidth]{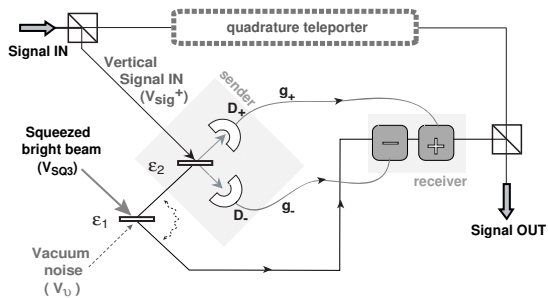}
       \caption{Biased entanglement teleporter experimental setup.
       $\varepsilon_{1}$: the variable transmittivity beamsplitter for
       biased entanglement of the inputs, $\varepsilon_{2}$: the 
variable
       transmittivity beamsplitter for detection, $D_{\pm}$: 
amplitude/
       phase homodyne detection, +/- : amplitude/ phase modulators,
       $g_{\pm}$: amplitude/ phase feedforward gains, PBS: polarising
       beamsplitter.}
     \label{BETelExp}
\end{figure}

It is somewhat disappointing that our SQD-teleporter scheme described
in section \ref{SQDtelSchm}, performs worse in terms of fidelity when
the SQ3 beam is squeezed (FIG.~\ref{SQDFidel}(b) and (c)).  In this
section we present an alternative polarisation teleportation scheme
that also uses three squeezed beams but can perform better than the
SQD-teleporter scheme in terms of fidelity.

Here, we use the third squeezed beam to generate entanglement of the
vertical polarisation.  A single squeezed beam and a vacuum mode are
combined on a beamsplitter (labelled $\varepsilon_{1}$), the outputs
of this beamsplitter then exhibit {\it biased entanglement}
\cite{Biased}.  That is, strong correlations are evident between the
squeezed quadratures of the two outputs, but only shot noise limited
correlations exist between the beams for the orthogonal quadratures
\cite{Biased}.  One of the biased entangled beams is then mixed with
the vertical mode of the input state at the second beamsplitter
(labelled $\varepsilon_{2}$).  The ability to choose the
transmittivity of this beamsplitter allows measurements of the
amplitude quadrature of the vertical signal, which is equivalent to
$\delta S_{1}$, or the phase quadrature of the vertical input, which
is not represented on the Poincar\'{e} sphere, or to alternatively
measure any combination of the two.  The signal is then detected and
fed-forward to the modulators.  We term this configuration biased
entanglement teleportation (BET) (see FIG.~\ref{BETelExp}).

The BET scheme can be thought of a modification of the twin
teleporters which tries to limit the resources from four bright,
squeezed beams, to only three.  One EPR pair is still maximally
entangled and teleports the horizontal fluctuations as before, however
the vertical information on the signal is teleported with one of the
squeezed beams turned off.  Further, the SQD-teleporter scheme from
Section \ref{SQDtelSchm} is a special case of the BET scheme and can
be recovered by setting $\varepsilon_{1}=$ 1 and $\varepsilon_{2}=$ 0.

The fidelity of a BET setup surpasses the $\mathcal{F}\ =\sqrt{(2/3)}$
direct detection limit.  To do this, various parameters of the system
can be optimised according to the value of squeezing injected,
$V^{\pm}_{\rm SQ3}$.  The beamsplitter transmittivities,
$\varepsilon_{1}$ and $\varepsilon_{2}$, can be changed to optimise
equation (\ref{fidelity}).  The amplitude modulator gain, $g_{+}$
which relates to the vertically polarised signal quadrature needs to
be kept at unity.  The phase modulator gain, $g_{-}$ however, relates
to the quadrature with no information, and hence is optimised to
minimise the reconstruction noise.  Both gains are functions of
$\varepsilon_{1}, \varepsilon_{2}$ and the squeezing, $V_{\rm
SQ3}^{\pm}$.  The polarity of $g_{+}$ and the quadrature being
squeezed (either $V^{-}_{\rm SQ3}$ or $V^{+}_{\rm SQ3}$), suggest four
possible operating regimes.  Our detailed analysis shows that three of
these regimes have maximum fidelity surpassing that of the
SQD-teleporter scheme.  For the remainder of this section we will only
discuss the best regime, which was obtained with feedforward gain of
$g_{+} >$ 0, and with the input phase squeezed ($V_{\rm SQ3}^{-} <$
1).

The improvement in the fidelity of the system occurs because at the
extremes of squeezing $\varepsilon_{1}$ and $\varepsilon_{2}
\rightarrow$ 0, so that almost all of the signal in the BET scheme is
directed to the amplitude detector, $(D_{+}$), while most of the phase
quadrature squeezing goes directly to the phase detector, $(D_{-})$. 
The modulation is then imprinted onto a nearly quantum noise limited
beam.  Some correlations exist between the detected phase fluctuations
and the fluctuations of the output beam, which enable a cancellation
of the output phase quadrature variance down to half the original shot
noise level.  The signal (amplitude) quadrature only pays the
measurement penalty by coupling to a single unit of vacuum noise.  For
identical squeezing levels on all three beams, $V_{\rm sq}^{+} <1$,
the expression for fidelity in terms of the beam splitter ratios is
given by
\begin{equation}
\mathcal{F}_{\rm max} = \frac{\
\mathcal{A}}{\sqrt{\mathcal{B} \mathcal{C}}} \label{FidelityBest} \\
\end{equation}
where $\mathcal{A}$, $\mathcal{B}$, and $\mathcal{C}$ are given by
\begin{widetext}
\begin{eqnarray}
\mathcal{A}&=&2 \sqrt{(\varepsilon_{2}-1) [\varepsilon_{2} (V_{\rm
SQ3}^{+}-1) (\varepsilon_{1}-1)-\varepsilon_{1} (V_{\rm 
SQ3}^{+}-1)-1]},
 \\
\mathcal{B}&=&2 \varepsilon_{2} (V_{\rm SQ3}^{+}-1)\
(\varepsilon_{1}-1)-\varepsilon_{1} (V_{\rm SQ3}^{+}-1) - 2,  \\
\mathcal{C}&=&\varepsilon_{2}\left(3-2\varepsilon_{1} +V_{\rm
SQ3}^{+}(2\varepsilon_{1}-1)\right) +2 (V_{\rm SQ3}^{+}-1)\
\sqrt{\varepsilon_{2}(1-\varepsilon_{2})\varepsilon_{1}(1-\varepsilon_{1})} 
\
-\varepsilon_{1}(V_{\rm SQ3}^{+}-1) - 3
\end{eqnarray}
\end{widetext}

\begin{figure}[ht]
     \centering
     \includegraphics[width=0.45\textwidth]{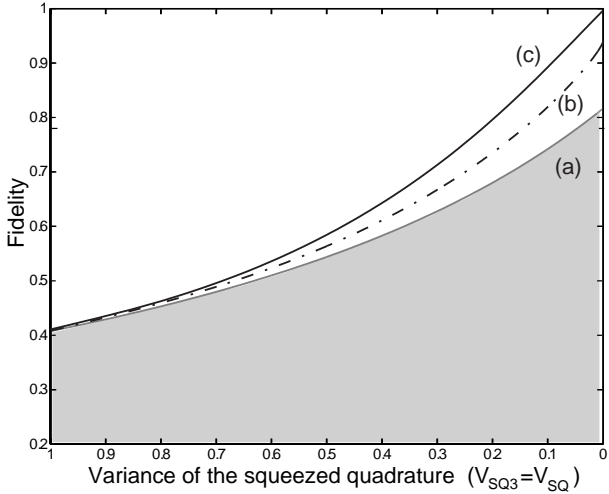}
     \caption{The comparison of the fidelity curves for (a)
     SQD-teleporter, (b) the optimum BET scheme, (c) the optimised
     twin teleporters.  The schemes are all equivalent at low
     squeezing parameter.}
     \label{BEFidel}
\end{figure}
FIG.~\ref{BEFidel}(b) illustrates the fidelity of the optimum BET
scheme by varying transmittivities, $\varepsilon_{1}$ and
$\varepsilon_{2}$, as a function of squeezing.  The maximum reached at
ideal squeezing is $\mathcal{F} = \frac{2\sqrt{2}}{3}\approx$ 0.943. 
As expected, unity fidelity is never reached, however for all input
squeezing levels the BET scheme is better than the SQD-teleporter
scheme.  Furthermore, the BET scheme can surpass the performance of
the twin teleporters scheme introduced in Section \ref{Twins} at
squeezing values within experimental reach.  This scheme preserves the
quantum nature of the complete state well but, as will be shown
shortly, when information transfer is considered it is inferior to
that of SQD-teleporter scheme.

%%%%%%%%%% TV %%%%%%%%%%%
The evaluation of the transfer coefficient and conditional variance is
also dependent on the optimisation of gain, beamsplitter ratios and
available squeezing parameters.  However the parameters optimised for
fidelity, do not necessarily correspond to the best T-V result.  This
occurs because fidelity weights every quadrature or Stokes operator
equally, whereas T-V analysis concentrates on the information
containing variables $S_{1}$, $S_{2}$ and $S_{3}$.  The BET system
then needs to be re-optimised and again, various regimes are reached
depending on the transmittivities of $\varepsilon_{1}$ and
$\varepsilon_{2}$.  Our analysis shows that $T_{V}^{+}$ and $V_{\rm
cv}$ as a function of feedforward gain are optimised if the BET
arrangement is set to recover the SQD-teleporter scheme of
FIG.~\ref{SQDTV}, by setting $\varepsilon_{1}=$ 1 and
$\varepsilon_{2}=$ 0.  Here the function $T^{+}_{V}=
(4g_{+}^{2})/({4g_{+}^{2} + V_{\rm sq}^{+}}) \rightarrow 1$ , as
$g_{+} \rightarrow \infty$.  With greater squeezing, the transfer
coefficient approaches unity more rapidly as $g_{+}$ increases.  The
amplitude quadrature conditional variance is independent of gain
$V_{\rm cv}^{+}=V_{\rm SQ3}^{+}$ and the minimum of zero occurs only
in the limit of perfect squeezing.

%%I had a bash at modifying this bit (ben 27/6/03)
It is clear from the above fidelity and T-V analysis that successful
information transfer is not necessarily linked to an improvement in
fidelity.  When optimising the fidelity, the BET protocol is weighted 
in favour of better state overlap.  This means improving the output 
phase noise of the SQ3 beam. Reducing this phase noise, however, 
means sacrificing signal and reducing the signal transfer 
coefficient. The decision of which characterisation method to use 
should be made dependent on the particular quantum information 
protocol, for which the teleportation scheme is to be used.

\section{Optimized twin teleporter scheme }

The fidelity curve as a function of squeezing for the twin teleporters
in the FIG.~\ref{SQDFidel}(a) could also be optimised for the
amplitude coded input signal considered in this paper.  This can be
achieved in a manner similar to the biased entanglement teleportation
optimisation, by adjusting the beamsplitter transmittivities for each
squeezing value.  When all four inputs are equally squeezed,
($V_{SQ3}=V_{SQ}=V_{\nu}$), and the pairs are 90 degrees out of phase
for best results, the fidelity is given by
\begin{equation}
    \mathcal{F}_{\rm 4SQ} = \frac{ \mathcal{D}}{\sqrt{\rm \mathcal{M}\
\mathcal{N}}} \label{4SQFidelityBest} \\
\end{equation}
where $\mathcal{D}$, $\mathcal{M}$, and $\mathcal{N}$ are given by
\begin{widetext}
\begin{eqnarray}
\mathcal{D}&=&2 \sqrt{V_{\rm SQ3}^{+} (\varepsilon_{2}-1)  
[\varepsilon_{2}\
(V_{\rm SQ3}^{+} -1)(V_{\rm SQ3}^{+} 
(\varepsilon_{1}-1)+\varepsilon_{1})\
+(V_{\rm SQ3}^{+})^{2}(1-\varepsilon_{1})+\varepsilon_{1}]},  \\ 
\nonumber
\mathcal{M}&=&(1+V_{\rm SQ3}^{+})\left( \varepsilon_{2} (V_{\rm
SQ3}^{+}-1)(1-2 \varepsilon_{1}) + \varepsilon_{1}(V_{\rm
SQ3}^{+}-1)-V_{\rm SQ3}^{+}\right),  \\ 
\mathcal{N}&=&\varepsilon_{2}\left(1-2 V_{\rm SQ3}^{+} - 2
\varepsilon_{1} -(V_{\rm SQ3}^{+})^{2}(1-2 \varepsilon_{1} )\right)  
\\  \nonumber
 &&+ (1-(V_{\rm SQ3}^{+})^{2})\left(\varepsilon_{1} -2
\sqrt{\varepsilon_{2}(1-\varepsilon_{2})\varepsilon_{1}(1-\varepsilon_{1})}\right)
+ 2 V_{\rm SQ3}^{+} + (V_{\rm SQ3}^{+})^{2} \nonumber
\end{eqnarray}
\end{widetext}
Again, several regimes emerge, however only the optimum regime for
fidelity is considered here.  This is shown on FIG.~\ref{BEFidel}(c).
The two optimised systems of BET (FIG.~\ref{BEFidel}(b)) and twin
teleporters, show comparable results at lower values of the squeezing
parameter, even though the twin teleporter requires more resources.

\section{Conclusion}

We have investigated schemes for the teleportation of polarisation
states carried by bright optical beams.  We have shown that simply
performing quadrature teleportation on the horizontal and vertical
constituent modes separately is not optimal in terms of squeezing
resources with respect to both the T-V and fidelity figures of merit. 
We introduce schemes that optimise the squeezing resources required
for polarisation teleportation with respect to each figure of merit. 
We find that the optimisation is different depending on the
figure of merit being used.  

The difference in optimisation of the two figures of merit can be
understood in the following way.  When small signals are applied to
the polarisation sidebands of a light field, they can be considered to
be a two-mode coherent state $|\alpha_{H} \rangle |\alpha_{V}
\rangle$.  Due to our choice of basis, both figures of merit quantify
the transfer of quantum information on the horizontal mode, however
they differ in how they treat the vertical mode.  The T-V analysis
considers the vertical mode to be a quantum limited classical 
channel. 
On the other hand the fidelity analysis considers the vertical mode to
carry quantum information on a restricted domain (i.e. $\alpha_{V}$ is
restricted to be real).  The appropriate figure of merit, and thus the
most efficient teleportation protocol to use in a particular
circumstance, depends on the way in which the quantum information is
being encoded.

\acknowledgments 

This work was supported by the Australian Research Council and is part
of the EU QIPC Project, No.  IST-1999-13071 (QUICOV).  We are grateful
to N.~Treps and H.~-A.~Bachor for useful discussion.
%%%%%%%%%%%%%%%%%%%%%%%%%%%%%%%%%%%%%%%%%%%%%%%%%%%%%%%%%%%%%%%

\end{document}

\end